\documentclass[12pt]{iopart}

\usepackage{citesort}
\usepackage{color}

\usepackage[margin=0.8in,top=0.8in,bottom=0.8in]{geometry}
\usepackage{setspace}
\usepackage{graphicx}

\begin{document}

\title{Thermalization of a two-dimensional photon gas in a polymeric host matrix}

\author{Julian Schmitt, Tobias Damm, Frank Vewinger, Martin Weitz, and Jan Klaers}

\address{\itshape Institut f\"ur Angewandte Physik, Universit\"at Bonn, Wegelerstra\ss e 8, D-53115 Bonn, Germany}

\begin{abstract}
We investigate thermodynamic properties of a two-dimensional photon gas confined by a dye-filled optical microcavity. A thermally equilibrated state of the photon gas is achieved by radiative coupling to a heat bath that is realized with dye molecules embedded in a polymer at room temperature. The chemical potential of the gas is freely adjustable. The optical microcavity consisting of two curved mirrors induces both a non-vanishing effective photon mass and a harmonic trapping potential for the photons. While previous experiments of our group have used liquid dye solutions, the measurements described here are based on dye molecules incorporated into a polymer host matrix. The solid state material allows a simplified operation of the experimental scheme. We describe studies of fluorescence properties of dye-doped polymers, and verify the applicability of Kennard-Stepanov theory in this system. In the future, dye-based solid state systems hold promise for the realization of single-mode light sources in thermal equilibrium based on Bose-Einstein condensation of photons, as well as for solar energy concentrators.

\end{abstract}


\maketitle
\newpage
\tableofcontents

\pagestyle{plain}

\section{Introduction}
Since the first experimental realization of Bose-Einstein condensation, much effort has been expended in further, extensive examination of this macroscopic quantum state of matter \cite{bose,einstein,ketterle,wieman}. This includes the extremely successful area of Bose-Einstein condensates realized with dilute, ultracold atomic gases \cite{sengstock}. More recently, condensation has also been reported with solid state quasiparticles, namely exciton-polaritons and magnons \cite{kasprzak,kasprzak2,yamamoto,balili,magnons}. Exciton-polaritons arise from bound electron-hole-pairs in semiconductors, which are strongly coupled to a light field within an optical resonator. Due to their short lifetime of typically $10^{-12}\ \textrm{s}$, a thermalization by polariton-polariton scattering processes is usually only observed at polariton densities above the condensation threshold \cite{kasprzak}. Two recent works using GaAs based microcavities have reported a crossover from an exciton-polariton to a photon-like lasing regime, the latter featuring a thermal wing coexisting with the lasing mode \cite{bajoni,kammann}. Both experiments have been carried out not far from the strong coupling regime, but nevertheless would be consistent with a thermalization mechanism that differs from interparticle scattering between the material parts of the coupled matter-light states. We here consider thermalization of a photon gas by means of repeated absorption-fluorescence processes of dye molecules embedded in a polymer within an optical microcavity in a regime of strong decoherence. Utilizing this fluorescence-induced thermalization mechanism, a Bose-Einstein condensation of pure photons in thermal equilibrium, both below and above the phase transition, could be recently observed in our group \cite{klaersnature}. Our earlier series of experiments was based on dye molecules in liquid solution \cite{klaersnature,klaersnaturephysics}.

The used dye molecules obey the Kennard-Stepanov law, which states that the frequency-dependent ratio of the absorption and fluorescence strength corresponds to a Boltzmann-function \cite{kennard,stepanov}. Such a relationship has been verified for dyes in liquid solutions, and also in the luminescence spectra of semiconductors and in vapor phase experiments \cite{merritt,ihara,moss,eastwood}. In our previous work, a liquid dye solution was placed between the mirrors of a high-finesse optical cavity. The cavity, with a mirror spacing in the micrometer regime, establishes a low-frequency cut-off for the photons in the optical regime and a non-vanishing effective photon mass. Inside the resonator, the photon dispersion acquires particle-like (quadratic) character, which along with an effective trapping potential induced by the curved mirrors makes the system formally equivalent to a two-dimensional gas of trapped, massive bosons. The photons thermalize to the temperature of the dye solution by repeated absorption and emission processes, fulfilling a detailed balance condition \cite{klaerstheorie}. In this system, we have observed a Bose-Einstein condensation of photons at room temperature \cite{klaersnature}.

These findings raise the question whether such an experimental scheme is also feasible in solid state environments rather than in liquid solutions promising an enhanced practicability of the setup. The occurrence of a photon thermalization is a priori uncertain, as the fluorescence characteristics of the dye molecules can be altered by changing solvents or host materials, including condensed matter substances in particular \cite{reisfeld,sah}. In liquid solution, it is well known that for many dyes the rovibrational sublevels of both the ground and electronically excited manifold are occupied according to a Boltzmann-distribution at the temperature of the solvent \cite{kennard,stepanov,sawicki}. This rovibrational equilibration is caused by frequent collisions between dye and solvent molecules, which occur many times (typically $\sim 1000$ at room temperature) during one absorption and fluorescence cycle of a molecule \cite{lakowicz}. In this way the equilibrated dye molecules imprint their thermalized state onto the photons simply by emission of fluorescence, as is discussed for the polymer-based system in greater detail in chapter \ref{therm}. Furthermore, we emphasize that the system under observation is in the weak coupling regime due to the collision-induced decoherence, which annihilates nearly all correlations between absorbed and emitted photonic states after a single absorption-emission process, as stated by Kasha's rule \cite{kasha}.

In the present work, dye molecules are embedded in an amorphous polymer at room temperature, which exhibits a high transmission in the visible spectrum. In such a host material rapid decoherence and vibronic equilibration on a femtosecond timescale is achieved by dissipation of vibrational energy of the molecules into the phonon bath of the polymer \cite{nakanishi}. We demonstrate the validity of the Kennard-Stepanov law for dye molecules in the polymer material by deriving the spectral temperature from the dye spectra. For this purpose, studies of absorption and emission spectra as well as the fluorescence quantum yield of the chromophores are carried out and compared to the spectroscopic properties of liquid dye solutions. Furthermore, we observe experimental evidence for a thermalized two-dimensional photon gas at the host temperature in the microcavity. According to the spectral distribution of the emitted radiation, the temperature of the photon gas is found to well agree with a room temperature spectrum above the low-frequency cavity cut-off. In addition, evidence for a thermalization of the photon gas is obtained from the observed spatial redistribution of fluorescence into the trap center, where the effective trapping potential induced by the mirror curvature exhibits a minimum value. At high average cavity photon numbers, a peak in the spatial intensity distribution is observed, which could be a hint to a Bose-Einstein-condensate. Due to fast and irreversible photobleaching of the dye molecules a more detailed investigation is precluded. Thus, no quantitative proof of a phase transition can presently be given.

In the following, chapter \ref{exp_scheme} describes the used experimental setup, while chapter \ref{char} presents measurements on the fluorescence properties of dye molecules embedded in polymer films. Further, chapter \ref{therm} contains measurements of the thermalized photon gas in the microcavity and discusses the fluorescence-induced thermalization process. Finally chapter \ref{concl} gives conclusions.

\section{Experimental scheme}
\label{exp_scheme}
\begin{figure}
\flushright	
		\includegraphics[width=13cm]{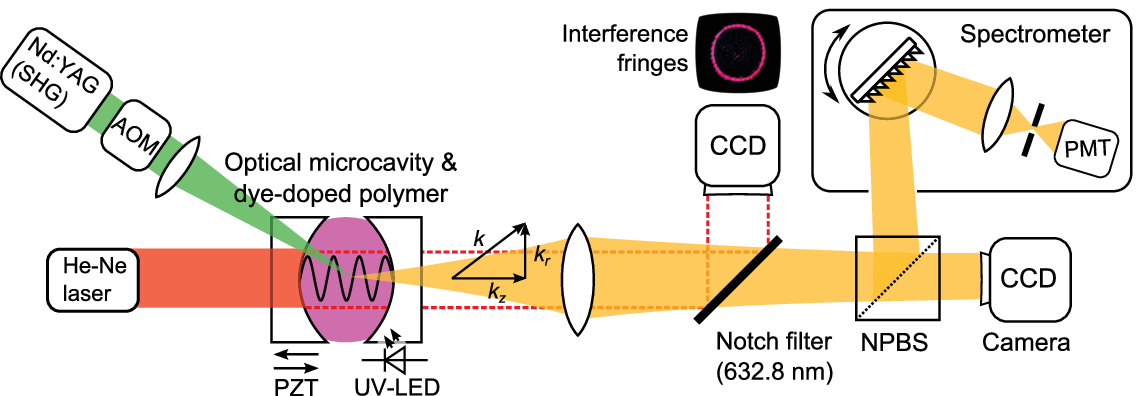}
\caption{\doublespacing Scheme of the polymer-filled microcavity experiment. Dye molecules are initially dissolved in a liquid photopolymer, which is inserted between the cavity mirrors, and subsequently cured by UV-LEDs. The mirror separation is adjusted to a value of $1.64\ \mu$m by piezo-tuning (PZT) and simultaneous monitoring of interference fringes produced by a He-Ne laser. A frequency-doubled Nd:YAG-laser is used for optical pumping of the resonator, where the intensity can be controlled by an acousto-optical modulator (AOM). The emitted light is investigated spectrally using a sensitive spectrometer based on a rotating grating and a photomultiplier tube (PMT), and spatially using a CCD camera.}
	\label{fig:setup}
\end{figure}

The used experimental setup, as shown in figure \ref{fig:setup}, consists of an optical microcavity filled with a dye-doped polymer, a pump source and both a charge-coupled device (CCD) as well as a spectrometer used for analysis of the emerging cavity light.

\begin{figure}
\flushright	
		\includegraphics[width=13cm]{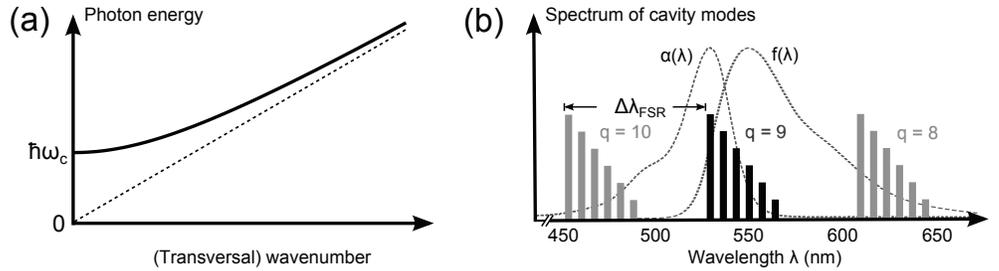}
\caption{\doublespacing (a) Modified dispersion relation of the cavity photons (solid) compared to free-space energy dispersion (dashed). The cavity cut-off $\hbar\omega_\textrm{\scriptsize c}$ is equivalent to the rest energy of the two-dimensional cavity photons. (b) Density of states of the microcavity. Preferential emission in the $(q=9,m,n)$-manifold constitutes a two-dimensional gas of photons. A linear increase in the mode density with energy accounts for the degeneracy of higher energy harmonic oscillator modes. Absorption $\alpha(\lambda)$ and fluorescence $f(\lambda)$ profiles (solute: R6G, solvent: ethanol) are shown for comparison (dashed).}
	\label{fig:disp_spectra}
\end{figure}

We ensure small photon loss rates by utilizing a high-finesse optical microcavity. It is formed by two Bragg mirrors (surface area $1\times 1$ mm$^2$) with a radius of curvature of $R=1$ m and a reflectivity above $99.997$\% in the spectral range from $500-590\ \textrm{nm}$, yielding a finesse of $\mathcal{F}\simeq10^5$ for the empty cavity. One mirror is mounted onto a voltage controlled piezoelectric crystal, allowing for an initial precise tuning of the mirror separation on the optical axis to the desired value of $D_0\approx 1.64\ \mu$m. This yields a free spectral range of two adjacent longitudinal modes $\Delta\lambda_\textrm{\scriptsize FSR} = \lambda^2/(2n_0 D_0) \approx 63$ nm, where $n_0=1.56$ is the index of refraction of the solid polymer medium inside the cavity at $570$ nm. This corresponds to a longitudinal wavenumber $q=9$, see figure \ref{fig:disp_spectra}(b).

The used dyes are either rhodamine 6G (R6G) or perylene diimide (PDI). In liquid solvents these dyes show a fluorescence quantum yield close to unity, $\Phi_{\textrm{\scriptsize R6G}} \simeq 0.95$ and $\Phi_\textrm{\scriptsize PDI}\simeq 0.97$ \cite{magde,wilson}. We assume that the rovibrational occupation of the molecular bands is thermally equilibrated by their contact to the phonon bath of the polymer host matrix, acting as a heat bath at room temperature. For a thermalized distribution of rovibrational levels both in the ground and electronically excited state of the dye molecule, it can be shown that by absorption and emission the photon gas thermalizes to the dye temperature \cite{klaerstheorie}. The thermalization process is discussed more precisely along with experimental results in chapter \ref{therm}.

The fluorescence bandwidth $\Delta\lambda_\textrm{\scriptsize dye}$ of both dyes is in the order of $80$ nm and therefore comparable to the free spectral range. We find that photons in good approximation are only emitted into transversal resonator modes TEM$_{9\textrm{\scriptsize mn}}$ with quantum numbers $(9,m,n)$, where the longitudinal degree of freedom is frozen out. On the one hand, excitation of photons into energetically higher modes with $q'=10$ is suppressed due to insufficient bandwidth of the dye. On the other hand, we do not observe considerable emission into highly excited transversal modes TEM$_{8\textrm{\scriptsize m'n'}}$ with the same energy as TEM$_{9\textrm{\scriptsize mn}}$ modes. We attribute this to their higher mode volume and consequently smaller overlap with the emitting molecular dipoles, which are predominantly located in the center of the resonator (see figure \ref{fig:disp_spectra}). Hence, fluorescence into modes with $q=9$ is intrinsically preferred by our system.

The cavity dispersion relation as a function of the transverse momentum in the paraxial approximation ($r \ll R, k_r\ll k_z(r)=q\pi /D(r)$) reads
\begin{equation}
E(k_r) = \frac{\hbar c}{n_0} \sqrt{k_r^2+k_z^2} \stackrel{k_r\ll k_z}{\simeq} \frac{\hbar^2 k_r^2}{2m_\textrm{\scriptsize ph}}+\frac12 m_\textrm{\scriptsize ph}\Omega^2 r^2+m_\textrm{\scriptsize ph}\left(\frac{c}{n_0}\right)^2,
\label{eq:dispersionrelation}
\end{equation}
where $m_\textrm{\scriptsize ph} = \hbar q \pi n_0/(c D_0)\simeq 9.5\times 10^{-36}$ kg is the effective mass of the two-dimensional photons and $D(r)$ the mirror separation at a distance $r$ from the optical axis. The harmonic oscillator trapping frequency $\Omega/2\pi = c/\sqrt{2\pi^2 n_0^2 D_0 R}\simeq 3.3\times 10^{10}$ Hz is specified by the curvature of the mirrors, and the rest energy of the cavity photons in frequency units is equivalent to the cut-off frequency $\omega_\textrm{\scriptsize c}/2\pi=m_\textrm{\scriptsize ph} c^2/(2\pi\hbar n_0^2) \simeq 5.3\times 10^{14}$ Hz (figure \ref{fig:disp_spectra}(a)). The transversal excitation energy above the cut-off is $u_{n,m}=(n+m)\hbar\Omega$ and the degeneracy of the resonator modes reads $g(u_{n,m}) = 2(u_{n,m}/(\hbar\Omega)+1)$, where the factor $2$ takes into account the two possible polarizations of each photon. Thermally equilibrated photons are expected to populate the energy levels in figure \ref{fig:disp_spectra}(b) according to a Bose-Einstein-distribution.

In our experiments, a commercially available polymer substance is applied as a host matrix for the dye molecules. The chromophores are solved in the optical adhesive NOA 61 (Norland Products), which is a colorless, liquid photopolymer curable by exposure to UV light. It offers a high optical transmission in the visible and near-infrared region. The liquid solution is purified using filters with $0.45\ \mu$m pore size to minimize scattering centers in the host matrix. After insertion of the viscous, liquid polymer the required mirror separation is adjusted by shining a He-Ne laser beam into the cavity. The reflectivity of the cavity mirrors at the He-Ne laser wavelength of $632.8\ \textrm{nm}$ is around $80$\%, and in transmission we observe circular interference patterns with radii depending on the distance between the mirrors, as indicated in figure \ref{fig:setup}. The red laser light is deflected from the optical axis using a notch filter and imaged onto a CCD camera. After tuning the cavity length to the desired cavity cut-off by adjusting to a certain diameter of the He-Ne interference rings, the thin polymer layer is cured by irradiation of light emitting diodes at $365\ \textrm{nm}$, while the cavity length is kept at a given value by maintaining the He-Ne laser interference pattern constant by piezo tuning. An adjustment of the cavity cut-off by using dye fluorescence proved to be unfeasible as the pump radiation leads to heat deposition in the polymer due to non-radiative decays. This causes structural defects by partial curing of the initially liquid polymer, which is avoided when using off-resonant He-Ne radiation.

Insertion of the initial photon gas and compensation of losses is achieved by optically pumping the dye-filled cavity with a beam derived from a frequency-doubled Nd:YAG laser near $532\ \textrm{nm}$ under an angle of $45^\circ$, exploiting the first maximum in the mirror transmission. At this wavelength, both dyes exhibit a high absorption cross-section (R6G: $\sigma(\lambda=532\ \textrm{nm}) = 4.4\times 10^{-16}$ cm$^2$; PDI: $\sigma(532\ \textrm{nm}) = 2.3\times 10^{-16}$ cm$^2$) \cite{du}. For thermalization measurements, we choose pump powers around $P_\textrm{\scriptsize pump} = 5\ \textrm{mW}$ (continuous wave). According to Beer's law, the fraction of pump power deposited in the cavity is approximately $P_\textrm{\scriptsize abs}/P_{\textrm{\scriptsize pump}}\propto T_{\textrm{\scriptsize 45}^\circ}\times(1-\exp\left[-\rho N_\textrm{\scriptsize A} \sigma(\lambda) D_\textrm{\scriptsize dye}\right])\approx 0.6$\% for typical dye concentrations of $\rho= 1\times 10^{-3}$ M. This calculation takes into account both the finite mirror transmission $T_{45^\circ}$ and also the effective absorption length in the dye-polymer-film $D_\textrm{\scriptsize dye}$. The latter deviates from $D_0$ due to a penetration of the light field into the mirror material by $q_0 = 4.68\pm 0.17$ halfwaves \cite{klaersapb}.

To avoid excitation of long-lived triplet states for experiments carried out at higher pump rates ($P_\textrm{\scriptsize pump} \approx 1 - 2\ \textrm{W}$), the incident light is chopped by an acousto-optical modulator into $\tau_\textrm{\scriptsize pulse}=500\ \textrm{ns}$ pulses with a repetition rate of $f_\textrm{\scriptsize rep}=250\ \textrm{Hz}$. The emitted cavity light is imaged onto a CCD chip and analyzed by a spectrometer, providing both spectral and spatial information on the trapped photon gas.

\section{Properties of dye molecules embedded in polymer films}
\label{char}
First we investigate the radiative properties of dye molecules incorporated into an amorphous polymer at room temperature. Therefore, the free-space absorption and fluorescence spectra of rhodamine and perylene diimide are recorded for a series of samples of different concentrations $\rho=(0.05 - 2)\times 10^{-3}$ M. The dye-doped photopolymer films of $0.65\ \textrm{mm}$ thickness were prepared between glass slides using precise spacers and subsequent UV-curing. Fluorescence is measured in the backward direction in order to minimize the effect of reabsorption within the sample. To record absorption spectra we have employed a transmission measurement using a broadband light source.

\begin{figure}[htbp]
\flushright	
		\includegraphics[width=12.7cm]{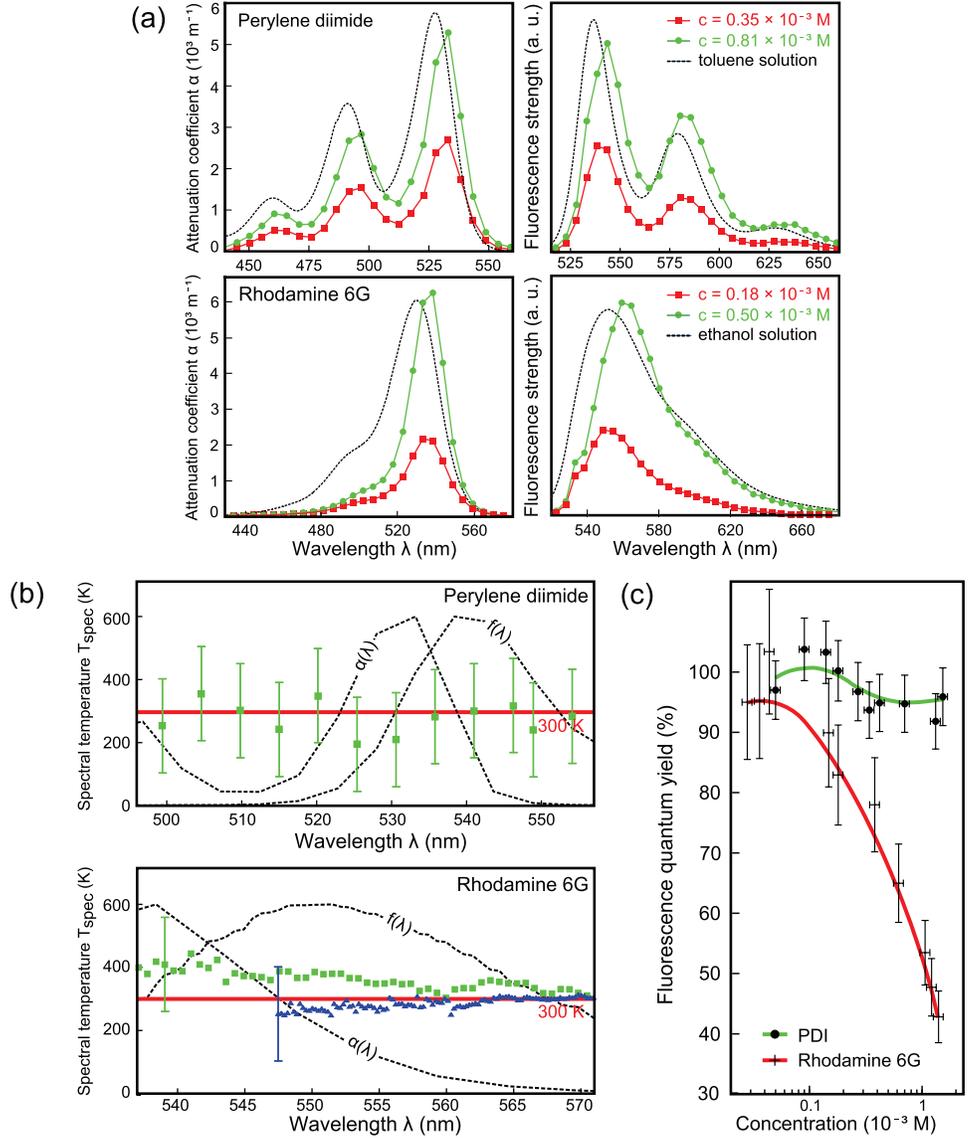}
	\caption{\doublespacing Optical properties of the dye-doped photopolymer for PDI (upper panels) and R6G (lower): (a) Absorption and fluorescence spectra, $\alpha(\lambda)$ and $f(\lambda$), in the cured photopolymer at room temperature with an excitation wavelength of $532$ nm for different concentrations. The global red shift of the emission and absorption lines are most likely due to a change of the solvent polarity compared to the liquid solutions (dashed line) \cite{iinuma}. (b) The spectral temperature (green) as calculated by equation (\ref{eq:spectraltemperature}). For guidance, $\alpha(\lambda)$ and $f(\lambda$) are added (dashed lines) and the error $\Delta T_\textrm{\scriptsize spec}=\pm 150$ K representative for all values is indicated. For both dyes $T_\textrm{\scriptsize spec}$ shows fair agreement with $300$ K. For comparison, measurements in liquid rhodamine solutions are shown (blue triangles). (c) Measured fluorescence quantum yield as a function of the dye concentration. For increased R6G concentrations a decay in quantum efficiency is observed (red line, crosses). PDI remains mostly unaffected by a variation of the dye concentration (green line, black dots).}
	\label{fig:spectra}
\end{figure}

Corresponding experimental results for concentrations $\rho \simeq (0.2 - 0.8)\times 10^{-3}$ M are shown in figure \ref{fig:spectra}(a) for both PDI (top panels) and R6G dye (bottom panels). For comparison, the dashed line gives results for spectra obtained in prior measurements with liquid solutions \cite{klaersapb}. The spectra of both dyes are found to be very similar to the corresponding liquid dye data, including the rovibrational substructure and the observed Stokes' shift. We note that both the absorption and the fluorescence profiles of both dyes in the solid state polymer environment exhibit a global redshift of $7.5\ \textrm{nm}$ with respect to the liquid solution data, which is a consequence of the modified interactions between the dye molecules and their environment. This result is in accordance to earlier reports on the spectroscopic behavior of R6G in polar and non-polar solvents, thin glasses and PMMA films \cite{reisfeld}. We find a zero-phonon line of $\lambda_{0} = 542\ \textrm{nm}$ for R6G and $\lambda_{0} = 534\ \textrm{nm}$ for PDI, respectively.

From the measured spectra we deduce the spectral temperature of the dye molecules as defined in Kennard-Stepanov theory \cite{sawicki},
\begin{equation}
T_\textrm{\scriptsize spec}=\frac{\hbar}{k_\textrm{\scriptsize B}}\left[\frac{\partial}{\partial \omega}\ln\left(\frac{\tilde\alpha(\omega)}{\tilde f(\omega)}\right)\right]^{-1}
\label{eq:spectraltemperature},
\end{equation}
where $\tilde\alpha(\omega)=\alpha(\omega)/\alpha(\omega_0)$ and $\tilde f(\omega)=f(\omega)\omega^{-3}/f(\omega_0)\omega_0^{-3}$ denote the dimensionless absorption and fluorescence profiles respectively in free space for an arbitrary $\omega_0$ \cite{klaersapb}. By repeated absorption and emission processes the photon gas is expected to thermalize to the spectral temperature, which usually corresponds to room temperature. However, deviations can arise if the fluorescence quantum yield is below unity \cite{klaersdis}. Figure \ref{fig:spectra}(b) shows the spectral temperature in the vicinity of the zero-phonon-line $\lambda_{0}$. For both dyes the determined spectral temperature is near the temperature of the dye-doped polymer of $300$ K, as expected from Kennard-Stepanov theory. The spectral temperature depends strongly on the derivatives $\partial\ln \alpha(\omega)/\partial \omega$ and $\partial\ln f(\omega)/\partial \omega$ of the experimental spectra, and the measurement accuracy of these quantities leads to an estimated uncertainty  of $\Delta T_\textrm{\scriptsize spec}=\pm 150$ K. For comparison, the blue triangles in figure \ref{fig:spectra}(b) show results for the spectral temperature of R6G in liquid dye solution at similar concentration ($\rho=10^{-3}\ \textrm{M}$).

Figure \ref{fig:spectra}(c) shows the measured quantum yields as a function of the dye-dopant concentration. It is determined by measuring the ratio between fluorescence and absorbed power of a dye-doped polymer sample, and by comparing it to a reference sample of known fluorescence quantum yield. As a quantum yield standard we used an optically thin liquid dye solution of R6G ($\rho=10^{-5}$ M, $\Phi_{\textrm{\scriptsize R6G}}\simeq 0.95$). While the quantum yield of PDI remains largely unaffected by a variation of concentration, a rapid decrease is observed for R6G. In contrast to this, fluorescence quenching in liquid rhodamine solutions becomes relevant only at concentrations above $10^{-2}$ M \cite{penzkofer}, indicating that the fluorescence properties are strongly affected by the combination of solvent and dye. The decreased fluorescence quantum yield in the R6G-doped polymer for larger dye concentrations is most likely caused by non-radiative energy dissipation. It can be explained by fluorescence resonant energy transfer (FRET), which takes into account the material-dependent relative orientation of the donor and acceptor dipoles \cite{foerster}. A similar behavior of fluorescence quenching of R6G in PMMA matrices was observed using dual beam thermal lens spectroscopy \cite{kurian}.

\section{Thermalization of the photon gas in the optical microcavity}
\label{therm}
Provided the previously discussed verification of the Kennard-Stepanov law in the dye-doped polymer, we have tested for thermalization of the photon gas in the dye-polymer-filled microcavity system in subsequent measurements. By insertion of the polymer film into a microresonator, the photonic mode density $g(u_{n,m})$ and, correspondingly, the fluorescence spectra of the dye molecules are modified, as described in chapter \ref{exp_scheme} in more detail. As a consequence, the photons occupy the transversal modes TEM$_{9mn}$ associated with a fixed longitudinal wavenumber $q=9$. If the mirror reflectivity is high enough to achieve reabsorption, a light-matter thermalization can be observed. The predicted mean occupation number at a temperature $T$ as a function of the transversal excitation energy $u_{n,m}$ reads
\begin{equation}
n_{T,\mu}(u_{n,m})=g(u_{n,m})\times f_\textrm{\scriptsize BE}(u_{n,m}) \stackrel{\mu\ll -k_\textrm{\tiny B}T}{\simeq}2\left(\frac{u_{n,m}}{\hbar\Omega}+1\right)\times e^{-\frac{u_{n,m}-\mu}{k_\textrm{\tiny{B}}T}},
\label{eq:occupationnumber}
\end{equation}
where $f_\textrm{\scriptsize BE}(u_{n,m})=\left(\exp[(u_{n,m}-\mu)/(k_\textrm{\scriptsize B}T)]-1\right)^{-1}$ denotes the Bose-Einstein distribution. The chemical potential $\mu$ is implicitly determined by $\sum_{n,m}{n_{T,\mu}(u_{n,m})}=\langle N\rangle$, with $\langle N\rangle$ as the average number of photons. These quantities are deducible both by measuring the cavity output power at a given cut-off energy and by taking into account the transmission of the cavity mirrors of $T\simeq 2.5\times 10^{-5}$. In our experiments, we typically measure $P\approx 5$ nW, corresponding to $\langle N\rangle \approx 8$ and $\mu\approx -9.1 \times k_\textrm{\scriptsize B} T$ (at $T=300$ K). By comparison, the expected critical photon number for a Bose-Einstein condensation is $N_{c} \simeq 1.1\times 10^5$. For small total photon numbers, $\mu\ll -k_\textrm{\scriptsize B}T$, the Bose-Einstein distribution becomes Boltzmann-like.

\begin{figure}
\flushright	
		\includegraphics[width=13cm]{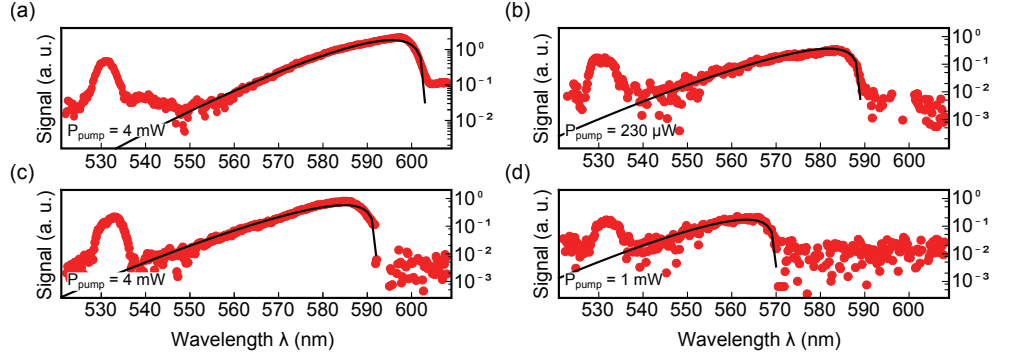}
	\caption{\doublespacing Spectral distribution of the cavity photons for different cut-off wavelengths. The data show good agreement to a Boltzmann distribution (according to equation (\ref{eq:occupationnumber})) at $T=300$ K and $\mu = -9.1\times k_\textrm{\scriptsize B}T$ (solid line). Residual pump light gives rise to a peak at $532$ nm. In (b) and (d) the signal-to-noise ratio decreases due to an increased photon reabsorption probability within the cavity at smaller cut-off wavelengths.\\
	(a),(b) R6G $\rho\simeq 0.7\times 10^{-3}$ M; (c),(d) PDI, $\rho\simeq 1\times 10^{-3}$ M}
	\label{fig:thermalisation}
\end{figure}

To experimentally verify the thermalization of the photon gas, we examined spectra of the emitted cavity light for various cut-off wavelengths ($\lambda_\textrm{\scriptsize c}=2\pi c/\omega_\textrm{\scriptsize c}=565-625\ \textrm{nm}$), pump powers ($0.23-4\ \textrm{mW}$) and different integration times. A series of spectra is shown in figure \ref{fig:thermalisation}. For the shown cut-off wavelengths between $569\ \textrm{nm}$ and $603\ \textrm{nm}$, a good agreement with the theoretical prediction of equation (\ref{eq:occupationnumber}) (solid line) is found in the region of $545$ nm to the respective cut-off wavelength. The thermal wing with its exponential decay to lower wavelengths, a characteristic signature for a Boltzmann distribution, is visible in all shown spectra, indicating the observation of a thermalized two-dimensional photon gas at room temperature. The peak at $532$ nm originates from residual pump light scattered at the resonator. Although the signal-to-noise ratio is reduced for smaller cut-off wavelengths, in the spectral regime under investigation evidence for a room temperature thermalization of the two-dimensional photon gas is found.

\begin{figure}
\flushright	
		\includegraphics[width=13cm]{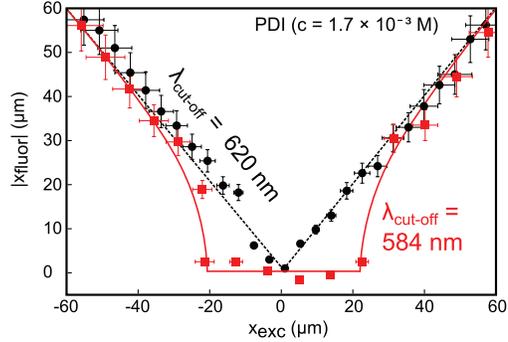}
	\caption{\doublespacing Spatial redistribution of fluorescence light for a displaced pump beam from the trap center, where the displacement is given by $x_\textrm{\scriptsize exc}$. The distance between the maximum of the fluorescence intensity and the minimum of the trapping potential is denoted by $|x_\textrm{\scriptsize fluor}|$. Upon variation of $x_\textrm{\scriptsize exc}$ the photons are spatially redistributed to the cavity center provided $\lambda_\textrm{\scriptsize cut-off}=584$ nm is in the reabsorbing regime fulfilling the thermalization condition (red squares). For $\lambda_\textrm{\scriptsize cut-off}=620$ nm, a coincidence of fluorescence and excitation spot is observed (black dots). This we attribute to a thermalization breakdown in the non-absorbing regime, where the photon gas does not exhibit a Boltzmann-distribution. This can also be seen in corresponding spectra, which are not shown. (PDI, $\rho= 1.7\times 10^{-3}$ M)}
	\label{fig:thermalisation2}
\end{figure}

In other measurements, we have investigated thermalization by varying the spatial position of the pump spot with respect to the center of the photon trapping potential. Figure \ref{fig:thermalisation2} shows the measured distance of the intensity maximum of the emitted radiation $|x_\textrm{\scriptsize fluor}|$ from the center versus position of the excitation spot $x_\textrm{\scriptsize exc}$. The red squares show data recorded with a cut-off wavelength near $584\ \textrm{nm}$. The fluorescence essentially freezes to the position of the trap center if the excitation spot is closer than approximately $20\ \mu\textrm{m}$ distance. For comparison, the black dots give data recorded when tuning the cut-off wavelength to $620\ \textrm{nm}$, where weak reabsorption prevents multiple scattering events and thus an efficient thermalization. The locking effect then breaks down. We note that the position locking effect can be seen as a direct consequence of the already spectrally observed thermalization, leading to a photon redistribution to the trap center, where the confining potential imposed by the curved mirrors exhibits a minimum value. The region where the spatial pulling into the center is observed is less than half the range as obtained in our earlier measurements carried out with liquid dye solutions \cite{klaersnaturephysics}. This could be due to larger photon losses from a reduced dye quantum efficiency in the polymer based system.

As inferred from the afore-mentioned measurements, there is a limitation in the extent of the spatial relaxation mechanism. This limitation is related to the average number of reabsorption cycles $\bar n_\textrm{\scriptsize re}$ a photon undergoes before it is lost from the cavity. For clarification, we will in the following provide an explanation on the thermalization procedure and give an estimate for the number of reabsorptions: After a typical free propagation time of $\tau_\textrm{\scriptsize ph}\approx 20\ \textrm{ps}$ in the dye-doped polymer-filled resonator, a photon is absorbed by a dye molecule. Reemission of a photon occurs after a molecular lifetime $\tau_\textrm{\scriptsize exc}$ of typically a few nanoseconds \cite{penzkofer}. The photon propagates freely again, and the absorption-reemission process is repeated. This type of thermalization constitutes a contact to a heat and particle reservoir and thus differs from a thermalization via particle-particle interactions as for cold atoms \cite{klaerstheorie}. The photon gas will be in a thermal equilibrium with the bath, if the rovibrational structure of the dye molecules is equilibrated before the reemission. This is easily fulfilled within the lifetime $\tau_\textrm{\scriptsize exc}$ of the dye, while the relaxation in the polymer host matrix is on subpicosecond timescales \cite{nakanishi}. We verified this by our measurement of the spectral temperature $T_\textrm{\scriptsize spec}\approx 300\ \textrm{K}$, see figure \ref{fig:spectra}(b).

In principle, only a single absorption-reemission cycle is required to thermalize the state of the photon in a spatially homogeneous system. This follows from Kasha's rule, stating that the fluorescence cycle is completely uncorrelated to the state of the absorbed light \cite{kasha}. However, the harmonic trapping potential constitutes an inhomogeneous system. If photons are generated far from the potential minimum, they first have to move into the center. While a spectral thermalization is achieved after a single absorption of a pump photon, a spatial relaxation towards the minimum of the trapping potential (from a displaced pump spot) requires a sufficiently high number of absorption-fluorescence cycles a photon undergoes before being lost from the system. Such photon losses can be due to non-radiative decay of the dye molecules (quantum yield $\Phi<1$), finite resonator finesse (mirror reflectivity $R<1$), as well as emission into modes not confined by the cavity. A measurement of the number of scattering events taking place within a lifetime of a photon inside the cavity has not yet been performed for the case of the polymer based host material. However, it is expected to be in the same order of magnitude for the solid state based setup and for the liquid dye scheme. We have estimated it based on a steady state condition by relating the intracavity power to the pumping power to be $\bar n_\textrm{\scriptsize re}\approx 3.8\pm 2.5$ for the liquid solvent case, as derived in an earlier article \cite{klaersapb}.

Compared to the vast number of collisions taking place in atomic gases, this value seems to be rather small. But one has to consider that nearly all correlations between the absorbed and emitted photon, except for a necessary spatial overlap between both photon states, have vanished after a single absorption-emission cycle \cite{kasha}. Therefore, the contact of the photon gas to the heat bath provides a stronger thermalization process compared to two-body collisions in atomic gases. We can experimentally verify the thermalization to operate properly both in the spectral and spatial regime, as investigated thoroughly in the case of the liquid dye system \cite{klaersnaturephysics}, see the data shown in figures \ref{fig:thermalisation} and \ref{fig:thermalisation2}, respectively, for the solid state based system discussed here.

\begin{figure}
\flushright	
		\includegraphics[width=13cm]{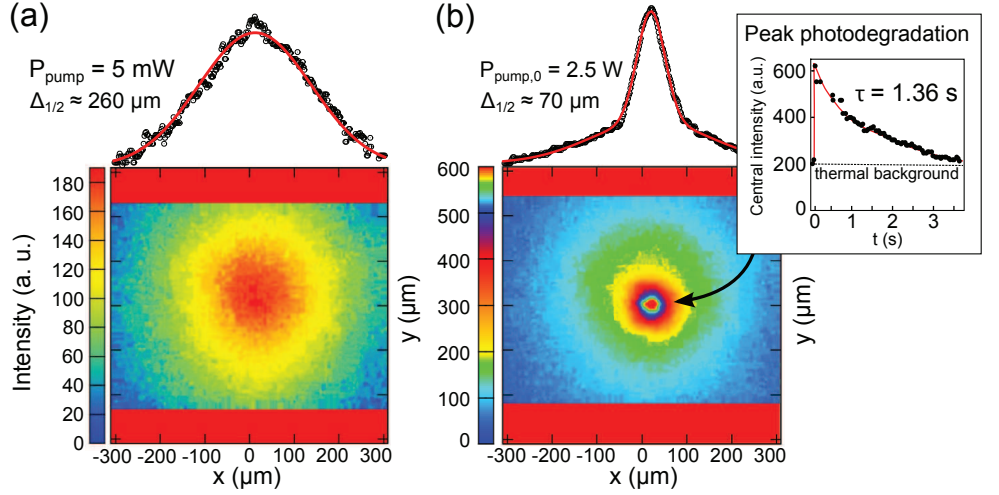}
	\caption{\doublespacing Images of the photon gas measured by a CCD-sensor (false-color image) and spatial intensity distributions along the $x$-axis. (a) For low average photon numbers inside the cavity, $\langle N\rangle\simeq8$ ($P_\textrm{\scriptsize pump}=5$ mW) the photon distribution is Gaussian. (b) A cusp in the center of the resonator plane emerges for higher photon numbers, $\langle N\rangle\approx 1.7\times 10^5$ ($P_{\textrm{\scriptsize pump}}=2.5\ \textrm{W}$, chopped into rectangular pulses with $\tau_\textrm{\scriptsize pulse}=500\ \textrm{ns}$, $f_\textrm{\scriptsize rep}=250\ \textrm{Hz}$). Rapid photobleaching ($\tau = 1.36$ s) of the dye molecules in the volume of the peak inhibits further investigation of this effect (inset, same intensity scale). After full photodegradation of the dye molecules in the trap center, the intensity distribution follows the shape of the thermal distribution in (a). (R6G, $\rho= 0.7\times 10^{-3}\ \textrm{M}$)}
	\label{fig:intensity}
\end{figure}

The spatial intensity distribution of the photons inside the resonator arises from a thermal average over all harmonic oscillator eigenfunctions $\Psi_{n,m}(x,y)$ and reads as
\begin{equation}
I_{T,\mu}(x,y) = \sum_{n,m\geq 0}{\frac{u_{n,m}+\hbar(\Omega+\omega_\textrm{\scriptsize c})}{\tau_\textrm{\scriptsize rt}}\left|\Psi_{n,m}(x,y)\right|^2 2 \left(e^{\frac{u_{n,m}-\mu}{k_\textrm{\tiny B}T}}-1\right)^{-1}},
\label{eq:intensity}
\end{equation}
where $\tau_\textrm{\scriptsize rt}=2D_0 n_0/c\simeq 17$ fs denotes the cavity round trip time for a photon. In our case ($T=300\ \textrm{K}$, $\mu=-9.1\times k_\textrm{\scriptsize B}T$) this corresponds to a Gaussian intensity distribution with a full width at half maximum of $230\ \mu$m. In figure \ref{fig:intensity} recorded CCD-images (false color) are shown for different pump powers. At low average photon numbers ($P_\textrm{\scriptsize pump}=5\ \textrm{mW}$) we observe a width of $\Delta_{1/2} = (260\pm 40)\ \mu$m for a Gaussian photon distribution, consistent with theoretical expectations. When increasing the pump power, the shape of the spatial distribution remains qualitatively unchanged up to a value of approximately $2.5\ \textrm{W}$. At pump powers above this value, we additionally observe a bright spot of light with a width of $\Delta_{1/2} = (70\pm 10)\ \mu$m emerging in the center of the spatial distribution (figure \ref{fig:intensity}(b)). By comparison, the diameter of the fundamental TEM$_{900}$ mode for an ideal Bose gas corresponds to $\Delta_{\textrm{\scriptsize ideal}} = 2\sqrt{\hbar \ln 2 /m_\textrm{\scriptsize ph}\Omega} \approx 12.2\ \mu$m. According to a previous observation of a Bose-Einstein condensation of photons in liquid dye solutions, the above described measurements suggest a similar behavior \cite{klaersnature}.

In the liquid-dye experiment the measured condensate width equals the diameter of the fundamental oscillator mode at moderate condensate fractions. Here we find a width increased by a factor of $\sim 6$ of the concentrated photon-distribution, which could be attributed to thermo-optical lensing caused by reduced quantum efficiency (as shown for R6G, figure \ref{fig:spectra}). In a more detailed analysis, the intensity-modulated index of refraction $n(\vec r)=n_0+\Delta n_r$, with $\Delta n_r = n_2 I(\vec r)$, can be interpreted as an effective photon-photon interaction in the Gross-Pietaevskii equation \cite{klaersapb}. If one considers the observed peak to be a broadened TEM$_{q00}$ mode, a dimensionless self-interaction constant $\tilde{g} \approx 2.7\times 10^{-2}$ would be necessary to be consistent with the measured diameter. For the liquid dye scheme, we have determined a dimensionless interaction constant of $\tilde{g}=7.5\times 10^{-4}$ \cite{klaersnature}. This is derived by comparing the measured diameter of an interacting Bose condensate in Thomas-Fermi-approximation to that of an ideal condensate,
\begin{eqnarray}{}
\Delta_\textrm{\scriptsize Thomas-Fermi} & = 2\sqrt{\hbar/(\sqrt{\pi}m_\textrm{\scriptsize ph}\Omega)}\left(\tilde g N_0\right)^{\frac14} = \Delta_\textrm{\scriptsize ideal} \left(\ln 2 \sqrt\pi\right)^{-\frac12}\left(\tilde g N_0\right)^\frac14
\label{eq:thomasfermi}
\end{eqnarray}
The condensate fraction is given by $N_0/N=1-N_c/N$ and from figure \ref{fig:intensity} it is estimated to $N_0/N=35\%$. Together with a total photon number of $N\approx 1.7\times 10^5$, which is obtained by integrating over the intensity distributions, we derive a ground state occupation of $N_0 \approx 6\times 10^4$. From this condensate fraction one deduces a critical photon number of $N_{c,\textrm{\scriptsize exp}} \approx 10^5$, which is consistent with the theoretical prediction mentioned above.

However, no detailed analysis of this effect, such as a spectral study, was carried out due to immediate photobleaching of the dye molecules within $\tau = 1.36\ \textrm{s}$ (inset of figure \ref{fig:intensity}), which turns out to irreversibly inhibit further fluorescence processes within the central volume of the cavity. Therefore, the values of $\tilde g \simeq 2.7\times 10^{-2}$, $N\approx 1.7\times 10^5$ and $N_{c,\textrm{\scriptsize exp}}\approx 10^5$ should be taken with care, since a progressed, partial bleaching of molecules in the moment of the data acquisition cannot be precluded. The intensity distribution after a full photodegradation of dye molecules in the cavity center is given by a Gaussian distribution with a peak-intensity $I(0,0)\simeq 200$ (in arbitrary units) as in figure \ref{fig:intensity}(a). In that case a repeated observation of a photon condensate using the same polymer film is impossible, but the cavity mirrors can be recycled by removing the layer from the mirror surfaces.

Different from the detailed verification of the photon Bose-Einstein condensate in liquid dye solutions, the presently described experiment lacks additional investigation of the critical particle number and the spectral distribution due to the short-lived chemical stability of the dye molecules embedded in the polymer. Photobleaching also occurs in liquid dye solutions \cite{beer}. However, due to constant molecular motion in liquid solvents this effect is less pronounced than in solid state systems with localized chromophores. For the polymer-based system a multi-mode occupancy cannot be a precluded by the discussed measurements. However, the results are evidently resembling a non-classical distribution of photons, which deviates from Boltzmann statistics. In the future, a more detailed investigation of the observed concentration of fluorescence light to the trap center will be necessary.

\section{Conclusion}
\label{concl}

We presented the observation of a thermalized, two-dimensional photon gas in a solid state system. A dye-doped amorphous photopolymer is incorporated into a high-finesse optical microcavity to allow a light-matter thermalization. An external pump source is employed to adjust the chemical potential of the photon gas and to compensate for photon losses. By investigating the emitted cavity light, we find that the spatial and spectral intensity distribution of the photons correspond well to that of a harmonically confined gas of particles at $300$ K. Furthermore, the observation of a thermalized photon gas is evidenced by the measurement of a fluorescence redistribution effect, which depends on the cavity cut-off.

The absorption and fluorescence profiles as well as the quantum yield of the used dye molecules embedded in the photopolymer-film are investigated and the spectral temperature is found to be near room temperature. The fluorescence quantum yield of R6G shows a decrease as a function of higher dye concentrations in the host matrix. However, at low R6G concentrations of around $0.5\times 10^{-3}$ M, fluorescence quenching mostly disappears and a thermalization of the photon gas is observed. In contrast to this, for PDI no significant dependency of the fluorescence quantum yield on the concentration of the dopant could be determined for the investigated dye concentrations. We observe a thermalization of photons to $300$ K in the PDI-filled microcavity for various concentrations. Further, at high average photon numbers a bright spot in the intensity distribution within the cavity is observed, which could be indicating a Bose-Einstein condensation of photons in the polymer-based microresonator system. The estimated critical photon number of approximately $10^5$ photons is of the same order as the theoretical expectation for the Bose-Einstein phase transition. Due to rapid photobleaching of the dye molecules, no further analysis could be carried out in the present work. 

Irreversible photodegradation becomes apparent as the main limiting factor for extended studies on the dye-polymer-samples at this stage. Therefore, we will prospectively study the incorporation of different dyes with high quantum yields into amorphous polymer matrices. The application of deoxygenated PMMA hosts with excellent optical quality, combined with photostable dyes are promising candidates for the described experimental scheme \cite{kurian,costela,rahn}. Altogether, photochemically stable setups hold promise for the realization of a new type of light source, which produces coherent light in thermal equilibrium. Additionally, the reported light concentration effect could be envisaged to enhance the efficiency of solar collectors \cite{vansark}. The usage of plastics rather than liquids as dye-solvents should enormously improve the practicability of such devices, including the possibility to use organic materials which can be pumped electrically. Other advantages are the mechanical stability and the protection of the mirrors due to the permanent polymer-based dye film between the cavity mirrors.

\section*{Acknowledgments}
We thank K Meerholz and D Hertel for useful discussions and collaboration on dye-doped PMMA thin-film-layers. Financial support from the Deutsche Forschungsgemeinschaft is acknowledged.

\section*{References}
\bibliography{thermalisation_polymer}

\newpage

\end{document}